\documentclass[
 floatfix,
aps,
prb,
twocolumn,
superscriptaddress,
amssymb,
]{revtex4}

\usepackage{float}
\usepackage{amssymb}
\usepackage{latexsym}
\usepackage[pdftex]{graphicx}
\usepackage{amsmath}
\usepackage{graphicx}
\usepackage{dcolumn}
\usepackage{amsfonts}
\usepackage{bm}

\newcommand{\be}{\begin{equation}}
\newcommand{\ee}{\end{equation}}
\newcommand{\bea}{\begin{eqnarray}}
\newcommand{\eea}{\end{eqnarray}}

\newcommand\fpk{\mbox{$f_{\mathrm{pk}}$}}

\def\sxx{\sigma_{\mathrm{xx}}}
\def\smax{\sigma_{\mathrm{pk}}}

\def\ec{E_{\mathrm{C}}}
\def\ez{E_{\mathrm{Z}}}

\def\ns{\mbox{$\nu^{\star}$}}

\def\tm{T_{\mathrm{m}}}
\def\to{t_{\mathrm{0}}}
\def\zo{Z_{\mathrm{0}}}
\def\pone{\Psi_{\mathrm{111}}}
\def\lb{\ell_{\mathrm{B}}}
\def\dsas{\Delta_{\mathrm{SAS}}}

\def\deg{^{\mathrm{o}}}

\begin{document}

\title{Wigner solids of wide quantum wells near Landau filling $\nu=1$
}

\author{A.\,T. Hatke}
\thanks{Present Address: Department of Physics and Astronomy, Purdue University, West Lafayette, Indiana 47907, USA}
\affiliation{National High Magnetic Field Laboratory, Tallahassee, Florida 32310, USA}

\author{Yang\,Liu}
\affiliation{Department of Electrical Engineering, Princeton University, Princeton, New Jersey 08544, USA}

\author{L.\,W. Engel}
\affiliation{National High Magnetic Field Laboratory, Tallahassee, Florida 32310, USA}

\author{L.\,N. Pfeiffer}
\affiliation{Department of Electrical Engineering, Princeton University, Princeton, New Jersey 08544, USA}

\author{K.\,W. West}
\affiliation{Department of Electrical Engineering, Princeton University, Princeton, New Jersey 08544, USA}

\author{K.\,W. Baldwin}
\affiliation{Department of Electrical Engineering, Princeton University, Princeton, New Jersey 08544, USA}
 
 \author{M. Shayegan}
\affiliation{Department of Electrical Engineering, Princeton University, Princeton, New Jersey 08544, USA}
\received{\today}

\begin{abstract}
Microwave spectroscopy  within the Landau filling ($\nu$) range of the integer quantum Hall effect (IQHE) has revealed pinning mode resonances signifying Wigner solids (WSs) composed of quasi-particles or  -holes. 
 We study pinning modes of WSs in wide quantum wells (WQWs) for $ 0.8\le\nu\le1.2$, varying  the density, $n$,    and tilting the sample by angle $\theta$ in the magnetic field.    Three distinct  WS phases are accessed.
  \ One phase,  S1, is phenomenologically the same as the WS observed in the IQHEs of narrow QWs.   The second phase,   S2, exists at $\nu$ further from $\nu=1$ than S1, and requires a sufficiently large  $n$ or $\theta$, implying  S2 is   stabilized by the Zeeman energy.\      The melting  temperatures of S1 and S2, estimated from the disappearance of the pinning mode, show different behavior  vs $\nu$.   
At the largest $n$ or $\theta$,  S2 disappears and the third phase, S1A,  replaces S1, also exhibiting a pinning mode. This occurs as  the WQW $\nu=1$ IQHE becomes a two-component, Halperin-Laughlin $\pone$ state.  We interpret S1A  as a  WS of the excitations of $\pone$, which has not been previously observed.
 
  

\end{abstract} 

\maketitle


\section{Introduction}
At the extremes of low temperature and high applied magnetic field, two-dimensional electron systems (2DESs) exhibit a vast array of phenomena.
Among the most prominent of these are the quantum Hall effects that can occur at integer (IQHE) and fractional (FQHE) values of the Landau level filling factor, $\nu=nh/eB_{\perp}$ where $n$ is the carrier density and $B_{\perp}=B_t\cos\theta$ is the perpendicular component of the magnetic field, and $B_t $ is the total magnetic field. 
At the  large $B_{\perp}$  termination of the FQHE series,   the kinetic energy of the  2DES is frozen out, and the preferred ground state is a Wigner solid (WS),\citep{lozovik:1975,lam:1984,andrei:1988,goldman:1990,jiang:1990,williams:1991,msreview,kunwc,archer:2013,rhim:2015,dengcommens} which  is  insulating due to pinning by residual disorder.  
In single-layer, n-type 2DESs, this termination occurs for $\nu\lesssim1/5$. 
Additionally, WSs have  been observed within  IQHEs by microwave spectroscopy,\citep{chen:2003,lewis:2004,zhu:2010b}  NMR,\citep{tiemann:2014}  and tunneling spectroscopy.\citep{jang:2017}    
IQHE WSs are composed of an effectively dilute concentration of quasi-particles or quasi-holes that crystallize in the presence of a filled Landau level within the $\nu$ range of IQHE plateau, though not exactly at integer $\nu$.

Microwave spectroscopy is of value for the study of WSs because these phases exhibit a characteristic microwave or rf resonance. \citep{andrei:1988,williams:1991,chen:2003,chen:2004,lewis:2004,zhu:2010,zhu:2010b,wang:2012,hatke:2014,hatke:2015,hatke:2017}
The  resonance is understood as a pinning mode, in which the solid oscillates within the disorder potential.  
This paper will describe the behavior of the pinning modes of multiple solid phases \cite{liu:2012,liu:2014c,hatke:2014}  found in the neighborhood of the IQHE around $\nu=1$ in    GaAs wide quantum wells (WQWs).    Pinning resonances  in narrower quantum wells, where only one subband is occupied, were observed within  IQHEs some time ago.\cite{chen:2003}   
The resonances are absent at exactly $\nu=1$,   but as $\nu$ moves away from $1$ in either direction they develop and  exhibit a maximum in intensity and sharpness, and disappear for $\nu$ farther than about $0.15$ from $1$.  The peak frequency, \fpk,  is largest for $\nu$ closest to $1$, and decreases monotonically as  $\nu$ moves away from $1$.  We refer to the solid that gives rise to the resonances seen in narrower wells  as S1.

In the WQWs that concern us in this paper, there is evidence from dc transport \cite{liu:2012,liu:2014c} and from earlier microwave spectroscopy\cite{hatke:2014} that an additional solid, which we call S2, can form under certain conditions on either side of $\nu=1$ at $\nu$ further away from $\nu=1$ than S1.  
Reference\,\onlinecite{liu:2014c} showed that  FQHEs for several $\nu$  on either side of $1$ exhibited spin state  transitions, which appeared at particular  Zeeman energies, $E_{\rm Z}$, at which the FQHE was suppressed.  
For $E_{\rm Z}$  sufficient to produce the FQHE transitions at $\nu=4/5$ or $6/5$ a reentrant integer quantum Hall effect  \cite{liu:2012} (RIQHE)  appeared around the fractional filling.  
The  RIQHE is a signature of insulating behavior in the partially filled Landau level.     The  $\nu$ range of the  RQIHE  is not  contiguous with the IQHE centered at $\nu=1$, but still has Hall resistance quantized at $h/e^2$ and vanishing dc longitudinal resistance. 
Reference\,\onlinecite{liu:2014c} ascribes the $4/5$ (or $6/5$) FQHE  in the WQW to a mixed state of different two-flux composite fermion\cite{jainbook} ($^2$CF) $\Lambda$-levels (analogous to Landau levels).
Here the lower $\Lambda$-level is spin up on both sides of the spin transition and the upper $\Lambda$-level changes from spin down to spin up.  
The emergence of the RIQHE was associated with a fully spin-polarized $^2$CF Wigner solid.\cite{archer:2011,archer:2013,rhim:2015}

The microwave spectroscopic studies on the same system\cite{hatke:2014} showed that at even higher density ($n$)  relative to the spin transition (hence higher $E_{\rm Z}$), a  microwave resonance was present well outside the normal $\nu$ range of the pinning modes of the IQHE centered at $\nu=1$, and at relatively higher peak frequency, $\fpk$.  
The $\nu$-region of enhanced \fpk\ was taken as a signature of S2  of which the RIQHE was a precursor seen in dc transport at lower  $n$  and  farther  from  $\nu=1$.
In all cases S2 gave way to  S1 as $\nu$ got  close enough  to $1$, and the pinning mode gave evidence of a solid-solid  transition from S2 to S1 as $\nu=1$ was approached.

In this paper we systematically study the evolution of the pinning mode resonance around $\nu=1$ in a WQW by investigating the role of $n$, {\em in situ} sample rotation, and temperature.
At relatively low $n$ and small  tilt angle ($\theta$) we observe a resonance that can be associated with  S1.
With increase of either $n$ or  $\theta$, (hence total magnetic field and  $E_{\rm Z}$)  S2 is observed as well, characterized by an enhanced-$\fpk$ region that occurs farther from $\nu=1$.  
Upon increasing $E_{\rm Z}$, starting from low $E_{\rm Z}$ without S2 present, S2 first sets in for $\nu<1$, at low $\nu$ (higher $E_{\rm Z}$, far from $\nu=1$), then sets in for $\nu>1$, again at low $\nu$ (in this case near $\nu=1$). 
This is consistent with an $E_{\rm Z}$ stabilizing S2, and we find that at all  $n$ and $\theta$,  S2 sets in for roughly constant  $E_{\rm Z}$ in units of the Coulomb energy. 
The temperature dependence is   different for the two solids, consistent with the two solids being made up of  different types of carriers such as  different flux-number CFs.    
Near the large $n$ and $\theta$ limits of our studies S2 again disappears; this occurs as the $\nu=1$ IQHE state becomes the  two-component  Halperin-Laughlin  $\pone$ state.  \citep{eisenstein:1992,murphy:1994,lay:1994,mueed:2016}  
The remaining   resonance, which is found close to $\nu=1$ like the S1 resonance,  is then a pinning mode of  WS of the quasi-particles or -holes of the $\pone$ state.   We denote this solid, for which the present paper provides the first evidence, as S1A.

\section{Experimental Methods}

\begin{figure}[t]
\includegraphics[width=2.5in]{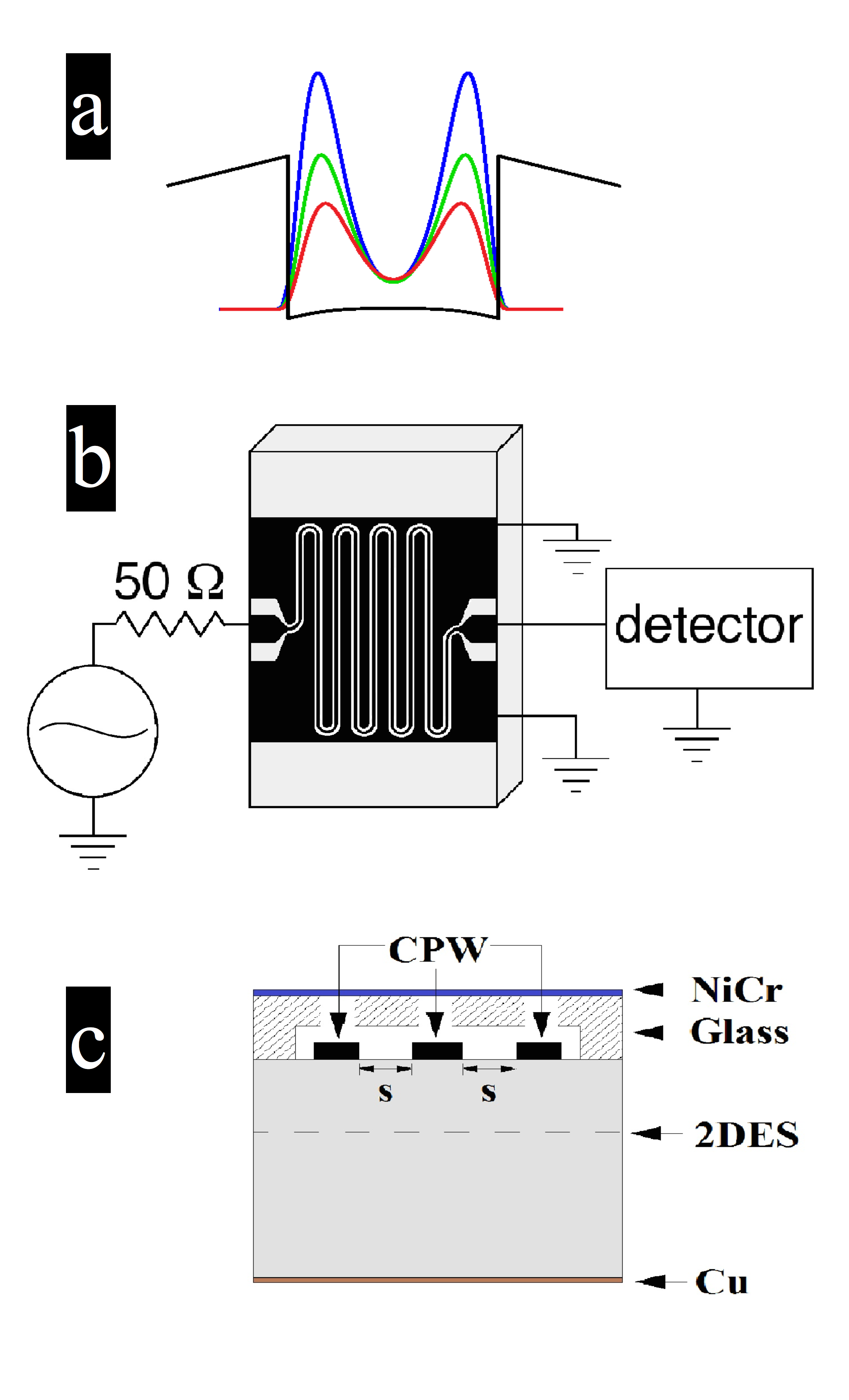}
\vspace{-0.2 in}
\caption{(Color Online)
(a) Charge distributions in a 65 nm wide quantum well, for densities $n=1.5$, $2.0$, and $2.5\times 10^{11}$ cm$^{-2}$.   
(b) Schematic of the microwave measurement set-up. The metal of the CPW is shown in black.
(c) The microwave set-up shown in a cutaway side view.
}
\label{cpw}
\end{figure}

Our microwave spectroscopy technique \cite{chen:2003,chen:2004,zhu:2010,wang:2012,hatke:2014,hatke:2015,hatke:2017}   uses a meandering coplanar waveguide (CPW) patterned in Cr:Au   on the sample surface.
Figure \,\ref{cpw}(b)    shows a schematic diagram of the microwave measurement technique and Fig.\,\ref{cpw}(c)  shows    a cutaway side view of the sample.
A Cu back gate was in direct contact with  the back of the sample and a NiCr front gate was deposited on a piece of glass that was etched to space it from the CPW by $\sim 10\,\mu$m, as in Refs.\,\onlinecite{hatke:2014,hatke:2015,hatke:2017}. 
We balanced the charge   in the growth direction  between the  front and back halves of the well   by biasing front and back gates such that individually each would change the carrier density by equal amounts.   
Charge asymmetry  of $10\%$ was found  not  to affect our results.
The samples were mounted on a rotatable sample stage with flexible, low-reflection broadband microwave microstrip in a dilution refrigerator running at $ 40\,$mK, unless otherwise noted.  
We find $\theta$, defined as the angle between the sample normal and applied magnetic field,  from the total magnetic field, $B_t$, of prominent IQHE features of known $\nu=nh/eB_\perp$.
The microwave measurements were carried out in the low-power limit, where the measurement is not sensitive to the excitation power.   

As in earlier work \citep{chen:2003,chen:2004,wang:2012,hatke:2014,hatke:2015} we calculate the diagonal conductivity as $ \sigma_{xx} (f) = (s/ l \zo) \ln (t/\to)$, where $s=30\ \mu$m is the distance between the center conductor and ground plane, $l=28\,$mm is the length of the CPW, $\zo=50\,\Omega$ is the characteristic impedance without the 2DES, and $t/\to$ is the normalized transmitted signal with $t$ the amplitude at the receiver and $\to$ the amplitude for $\nu=1$.  
Hence  $\sigma_{xx} (f)$ is the difference between the conductivity and that for $\nu=1$; just at $\nu=1$ the conductivity is vanishing at low temperature.  
Using $\to$ to normalize for $\nu=2$  instead produced a  shift of the spectra by $\lesssim 1 \mu$S, which does not affect our conclusions.    In this paper, the peak frequency of a resonance, $\fpk$, is found using the condition\citep{hatke:2014} Im$(\sigma_{xx})=0$.

To cover the density range of interest,  microwave spectra were obtained from two WQW  wafers, Sample A and Sample B,  with well width $w=65$ nm but as-cooled densities,  $n=1.4$ and $n=2.5$  and mobilities $\mu=5.2 \times 10^{6}\,$cm$^{2}/$Vs and $8.8  \times 10^{6}\,$cm$^{2}/$Vs, respectively.  Throughout the paper $n$ will retain units of $10^{11}$ cm$^{-2}$, which will be omitted for brevity.   Sample A could only be gated up to 1.97 and  Sample B could only be gated down to 2.28.  
Figure\,\ref{cpw}(a) shows charge distributions  for a 65 nm WQW with densities like those studied here.

\section{Results}
\subsection{Spectra as density and tilt angle change}

This section will  show the development of the experimental signature of S2, which 
appears as $n$ or $\theta$ increase, but vanishes at the largest $n$ or $\theta$.

\begin{figure*}[t]
\includegraphics[width=.75\textwidth]{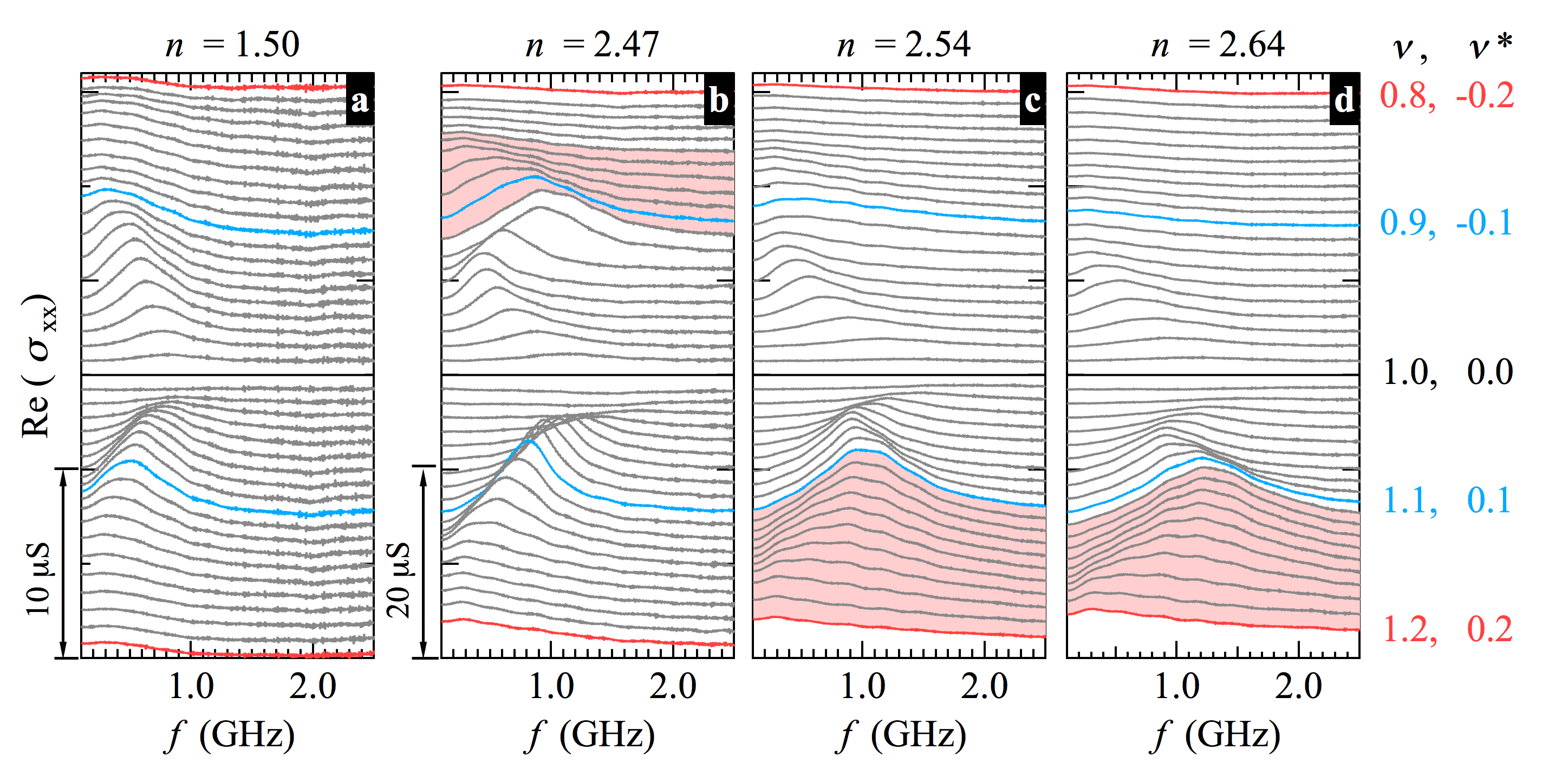}
\vspace{-0.2 in}
\caption{(Color Online)
Microwave spectra, Re\,$(\sigma_{xx})$ vs $f$, at fixed $\nu$ from $0.8$ (top) to $1.2$ (bottom) with a step of $0.01$ at different $n$, as marked.
Both $\nu$ and $\ns=1-\nu$ are marked along the right axis.  
(a) Microwave spectra for Sample A, successive traces vertically offset from each other  by $0.75\, \mu$S. (b)-(d)  spectra for  Sample B, successive traces  vertically offset from each other by $1.5\, \mu$S. Shaded areas mark spectra from well-developed S2. 
}
\label{den}
 \end{figure*}
 \begin{figure}[b]
\includegraphics[width=2.5in]{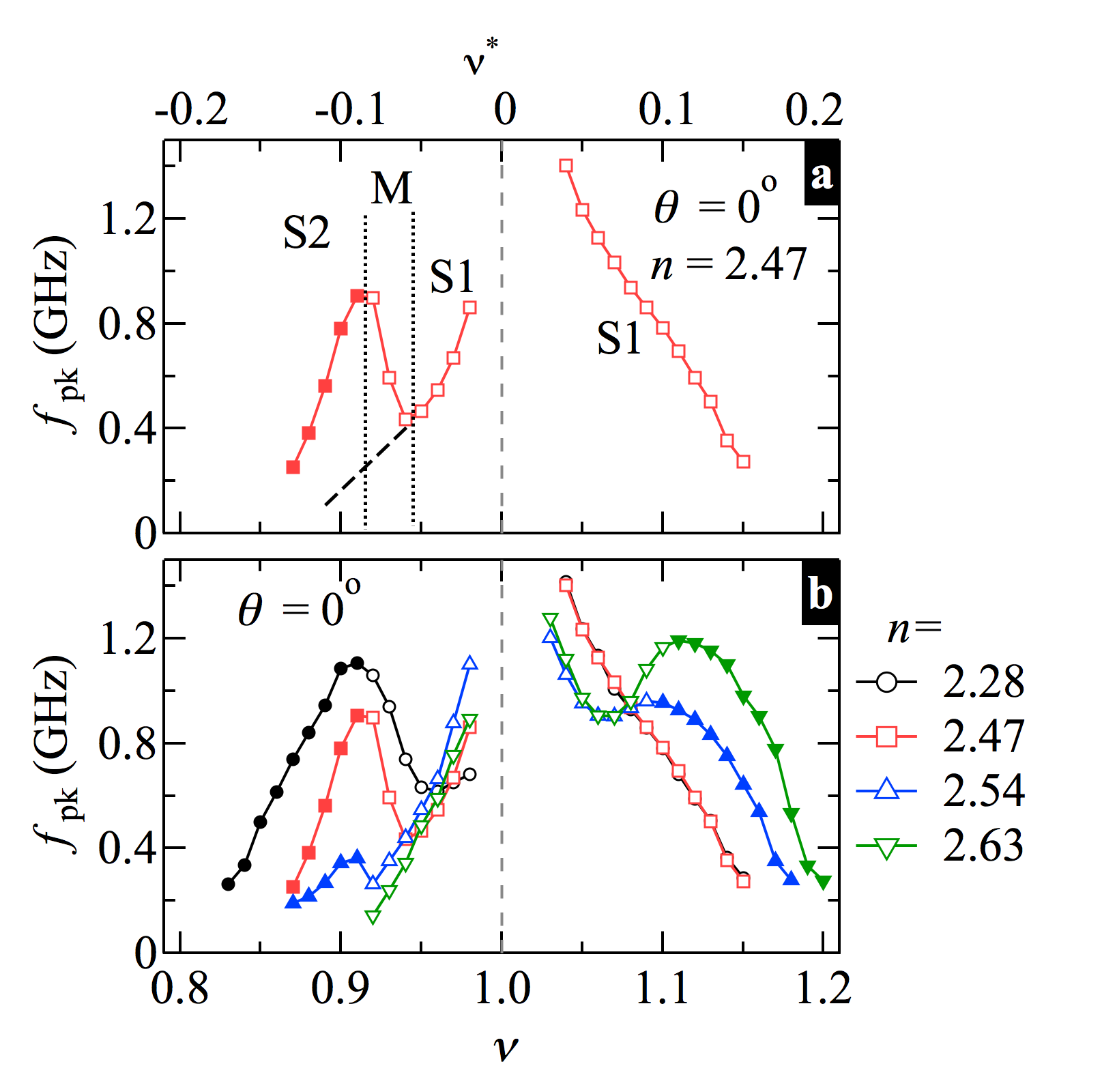}
\vspace{-0.2 in}
\caption{(Color Online) Data for Sample B. 
(a) $\fpk$ vs $\nu$ for $n=2.47$, tilt angle $\theta=0\deg$. The regions of solids S1, S2 are marked, as is the region (M) of 
transition between S1 and S2.  Symbols within well-developed S2 are filled.   
(b) $\fpk$ vs $\nu$ at different $n$ with sample perpendicular to magnetic field, $\theta=0\deg$. For $\nu>1$ the  data for $n=2.28$ lie directly under those for $n=2.47$. 
}
\label{fpkn}
\end{figure}
\begin{figure*}[t]
\includegraphics[width=.75\textwidth]{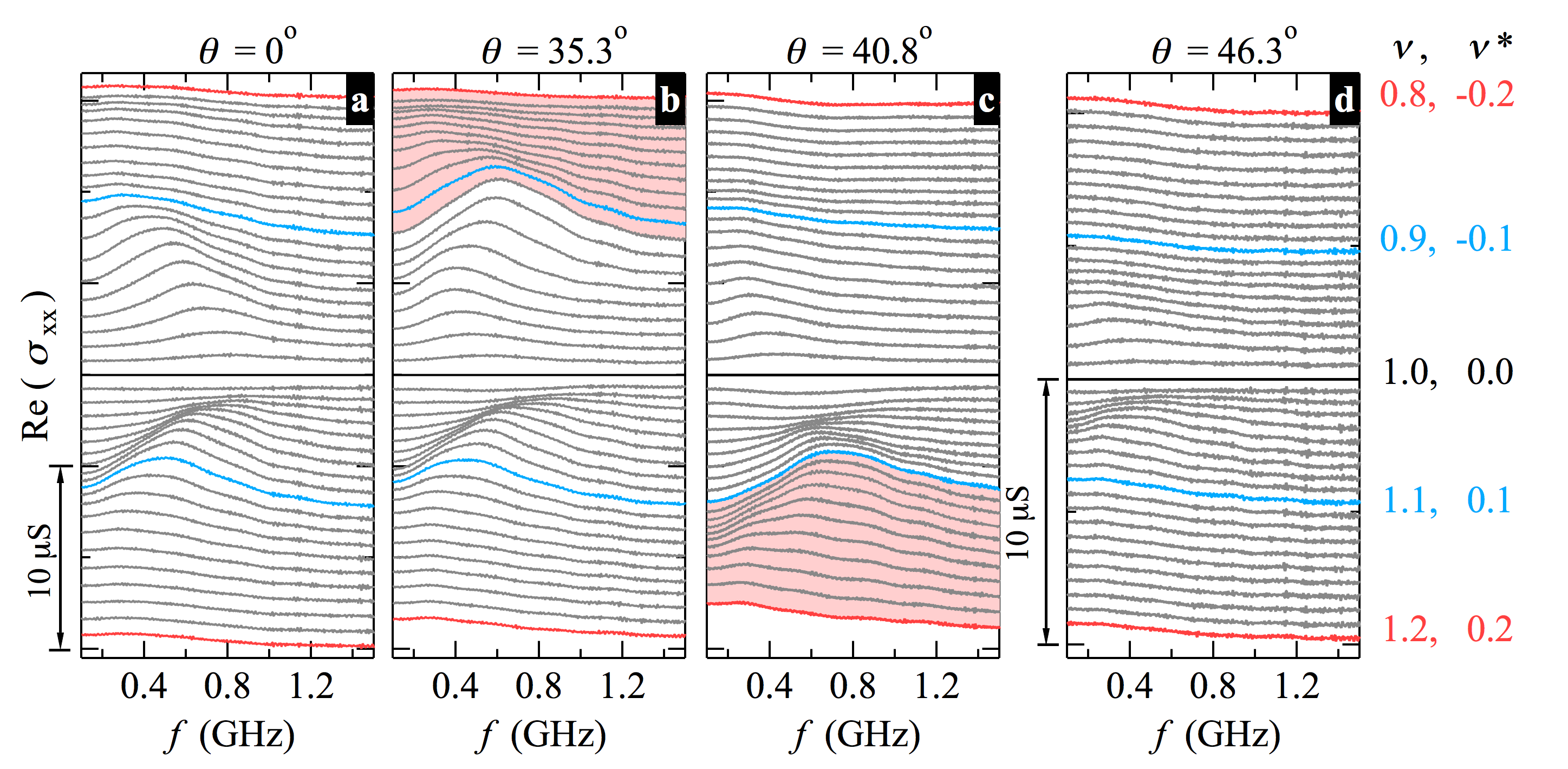}
\vspace{-0.2 in}
\caption{(Color Online)
(a)-(d) Microwave spectra, Re\,$(\sigma_{xx})$ vs $f$, at fixed $\nu$ from $0.8$ (top) to $1.2$ (bottom) with a step of $0.01$ for $n=1.50$ at different tilt angles from Sample A.
Both $\nu$ and $\ns$ are marked along the right axis of (d). Successive traces are vertically offset by $0.75\,\mu$S in  (a)-(c) and by $0.5\,\mu$S in (d). Traces with heavier lines correspond to the marked $\nu$. Shaded areas mark spectra from well-developed S2.
}
\label{tilt}
 \end{figure*}
 \begin{figure}[b]
\includegraphics[width=2in]{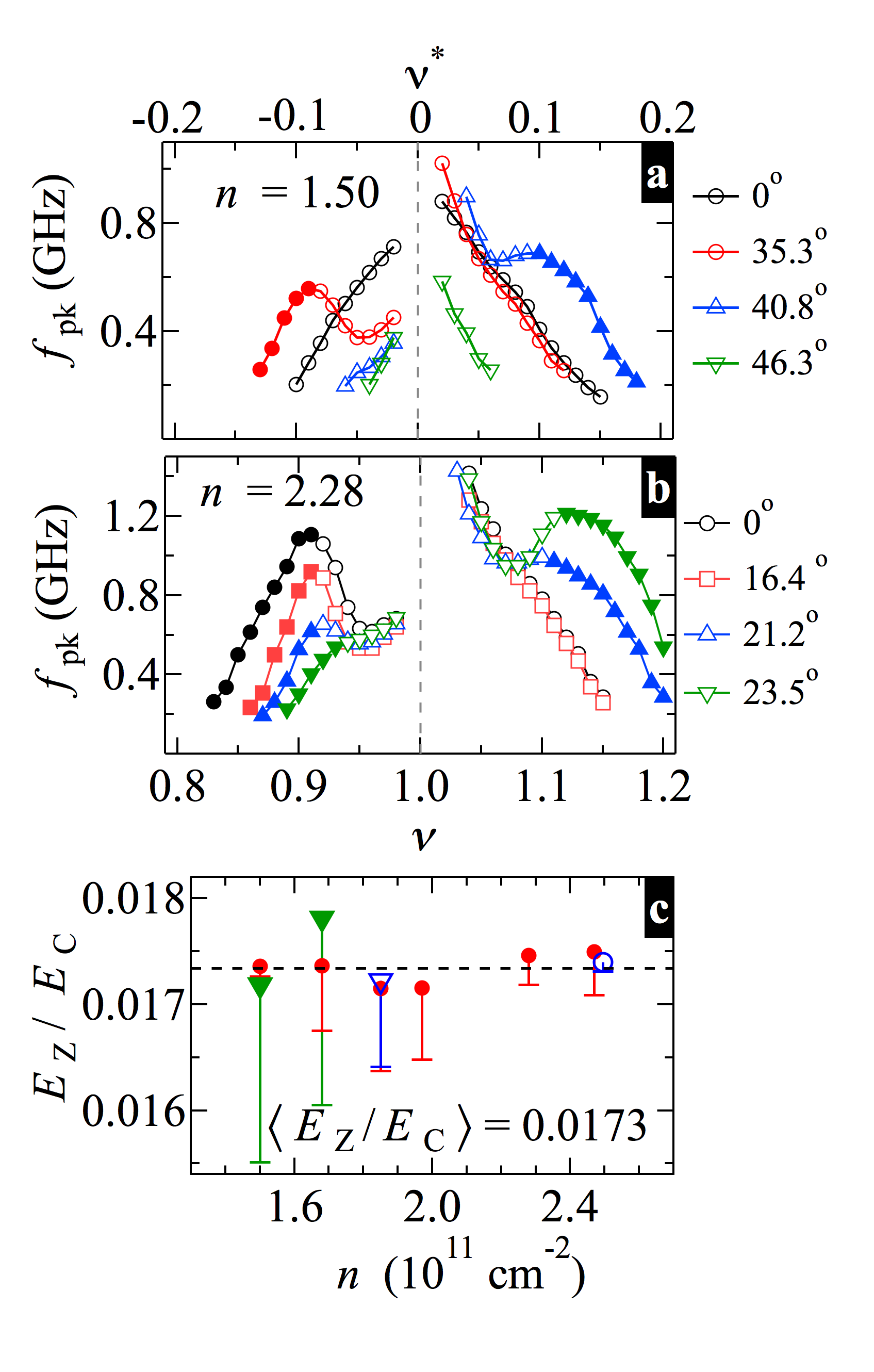}
 \vspace{-0.2in}
\caption{(Color Online)
(a) $\fpk$ vs $\nu$ for $n=1.50$ at different $\theta$, for Sample A.   Symbols within well-developed S2 are filled.  
(b) $\fpk$ vs $\nu$ for $n=2.28$ at different $\theta$, for Sample B.   Symbols within well-developed S2 are filled.  
(c) Zeeman energy in units of Coulomb energy, $\ez/\ec$, at the lowest $\ez/\ec$ for  which the S2 resonance is observed (see text),  
 plotted vs density, $n$.  Triangles  are for $\nu<1$, circles  are for $\nu>1$. Closed symbols are obtained by varying $\theta$ at fixed $n$; open symbols are for $\theta=0$, and are estimated by varying $n$. The  dotted line marks the  average for the points, $\langle \ez/\ec=0.0173 \rangle$.    Data with  $n\ge2.28$   are from Sample B, lower $n$ data are from Sample A.}
\label{fpkt}
\end{figure}

Figures \,\ref{den}(a)-(d) show   Re\,$(\sigma_{xx})$ vs $f$ spectra, at different $n$, for pinning modes near $\nu=1$.  The spectra are offset upward proportional to $\nu$, which is marked at right.   
For reference, the development of S1 alone is shown in Fig. \,\ref{den}(a), for Sample A at $n=1.5$. 
 As $\nu$ moves away from 1, the resonance frequency decreases,  while the amplitude first increases then eventually vanishes.   We ascribe this behavior,  first observed\cite{chen:2003} long ago,  to the evolution of a single solid, which we call S1, as its carrier density changes.   Because the solid is made 
up of quasi-particles or -holes created as $\nu$ moves away from one, its charge density is proportional to $|\ns|$, where $\ns=\nu-1$.  
While in some cases the quasi-particle and -hole \fpk\ vs $|\ns|$ are nearly symmetric around $\nu=1$, this is not the case for much of the data presented here.  The lack of symmetry could plausibly  be due to the features of the disorder potential (which may for example have sharp spikes of one sign only)   not acting in the same way on positively and negatively  charged WS.  
The dependence on charge density, which we call the ``density effect,"  is generic to pinning modes,\cite{clidensity,chen:2003,wang:2012} and  
 is qualitatively predicted in weak-pinning theories.\cite{fertig:1999,chitra:2001,fogler:2000}
  WSs of lower  charge density (here, smaller $|\ns|$) have larger  $\fpk$ because they are less stiff, and carrier  positions are more closely associated with the disorder   potential.\cite{clidensity}   
The intensity of the resonance initially increases because the charge density of the solid increases, but at large enough $|\ns|$ falls off as the solid gives way to fractional quantum Hall liquid states.

Figure  \ref{den}(b) shows    spectra for  $n=2.47$  in Sample B. The development of spectra for $\nu>1$ is similar to that in Fig. \,\ref{den}(a), but  is markedly  different for $\nu<1$, where the signature of S2 develops.   This signature is most easily shown in a graph of \fpk\ vs $\nu$.    Figure \ref{fpkn}(a), shows such data from the spectra in Fig. \,\ref{den}(b)   with \ns\ also on the  top axis, to  summarize  the  interpretation\cite{hatke:2014} of the data in terms of S1 and S2. 
   At this $n$, for $\nu>1$, \fpk\ decreases  monotonically  as $|\ns|$ increases; this is interpreted as due to the density effect in a single solid (S1). 
    For $\nu<1$, as $|\ns|$  first increases, in the region marked S1, \fpk\ decreases, then shows a sharp upturn in the region marked M, then begins to decrease again.    The region of increase of  \fpk\  with   $|\ns|$   cannot be explained  by the density effect, so the existence of another WS phase is implied,\cite{hatke:2014} which we call S2.   The region marked S2  shows  \fpk\ again decreasing with  $|\ns|$,  because the density effect acts on S2.  The region M is  interpreted  as a transition,  with a mixture of S1 and S2.    The points in the S2 region are filled, as will be the convention for the rest of this paper.  The signature of S2 is the region of  \fpk\ enhanced above what would be expected by extrapolating S1 vs $\ns$.  The extrapolated \fpk\ for S1 is represented by a dashed line in the  Fig. \,\ref{fpkn}(a).

In Fig.  \,\ref{den}, spectra from S2 are highlighted by shading between the traces.    
While Fig.  \,\ref{den}(b) shows S2 only for $\nu<1$,    
  Figs.   \,\ref{den}(c) and (d), show that on increasing $n$ further, 
 the enhancement of \fpk\ in S2   becomes weaker  for $\nu<1$ as it  emerges and grows stronger for   $\nu>1$.
At $n$ of 2.54, in  Fig. \,\ref{den}(c),   S2 is present on both sides of $\nu=1$.  
  For Fig. \,\ref{den}(d),  with data for  $n=2.64$,   only S1 is present for $\nu<1$, although S2 remains well-developed for $\nu>1$.

The  \fpk\ vs $\nu$ curves  of  Fig. \ref{fpkn}(b)   summarize the development of S2 with increasing $n$.  
For $\nu<1$, the  \fpk\ enhancement of S2 is largest for  $n=2.28$, and is absent for $n=2.63$.  For $\nu>1$, S2 is not visible for $n<2.54$, and 
the amount of \fpk\ enhancement  due to S2 is increasing up to the largest $n$ of 2.63.

Increasing $\theta$ has much the same effect as increasing $n$,  causing 
 the signature of S2 to turn on, then subside.  This is illustrated as spectra in  Fig. \,\ref{tilt} and   as \fpk\ vs $\nu$ in Figs. \,\ref{fpkt}(a)      and (b).    In  Fig\, \ref{tilt}   the   spectra are from Sample A with $n=1.50$,  at different  $\theta$.  
  Figure\,\ref{tilt}(a), for $\theta=0$ (same as Fig.\,\ref{den}(a)),  shows no sign of S2.   Figure\,\ref{tilt}(b), with spectra taken at  $\theta=35.3\deg$, shows S2 for $\nu<1$ only, but is essentially the same as panel (a) for $\nu>1$.    In Fig.\,\ref{tilt}(c), with data taken  at larger $\theta=40.8\deg$, S2 has subsided for $\nu<1$, but is well-developed for $\nu>1$.
 Finally, in panel (d),  for $\theta=46.3\deg$ there is no signature of S2 on either side of $\nu=1$.  
  
   Figures \ref{fpkt}(a) and (b)    show the evolution of \fpk\ vs $\nu$ 
  with $\theta$.  In Fig.\,\ref{fpkt}(a), for Sample A at $n=1.50$,  as $\theta$ increases, S2 emerges and then disappears, first for $\nu<1$, then  for $\nu>1$.    
  In Fig.\,\ref{fpkt}(b) at $n=2.28$, for Sample B, the development with increasing $\theta$  is similar, but   S2 is already present at $\theta=0$.


The observed appearance of the S2 signature as either $n$ or $\theta$ are increased is consistent with S2 
becoming visible at sufficiently large total magnetic field $B_{t}$, which at fixed $\nu$ increases with both $\theta$ and $n$.  This suggests the Zeeman energy, $E_{\rm Z}=g\mu_B B_t$, where $|g|=0.44$, is  important to stabilizing  S2, as  Ref. \,\onlinecite{liu:2014c} pointed out for the RIQHE phases that lie further from $\nu=1$.  The {\em disappearance} of S2 at still higher $n$ or $\theta$ is dealt with below in Section III.C. 

To quantify the role of Zeeman energy we present the {\em minimum}   $\ez/\ec$ required to observe S2, where the Coulomb energy is defined as $\ec= e^{2}/4 \pi\epsilon_0\epsilon\lb$ and $\lb=(\hbar/eB_\perp)^{1/2}$ is the magnetic length.  
 For each $n$, we find the
minimum $\theta$ for which S2 can be seen, then calculate $\ez/\ec$ at the highest $\nu$ (lowest $B_t$) within S2. This gives a minimum  $\ez/\ec$ for each $n$, plotted in Fig. \ref{fpkt}(c) as closed symbols.  We do not continuously vary $\theta$, so the differences between $\theta$ for which S2 is seen and $\theta$ for which it is not are used to obtain error bars. In addition, for $\theta=0$ we measured $\fpk$ vs $\nu$ at many fixed $n$, as exemplified by the data in Fig. \ref{fpkn}(b). We also estimate a minimum   $\ez/\ec$ from the lowest $n$ for which S2 can be seen, choosing the lowest-$B_t$ point within S2.    Figure \ref{fpkt}(b) shows data for S2 minimum $\ez/\ec$ required to observe S2 for $\nu<1$ (triangles) as well as for $\nu>1$.  Averaging all the minimum $\ez/\ec$ obtained in this way gives  an estimated  $\ez/\ec$  of 0.017 for  the emergence of S2.  The close grouping of the points in Fig. \ref{fpkt}(c)  is consistent with $\ez$ stabilizing S2.

\begin{figure}[t]
\includegraphics[width=0.6\textwidth]{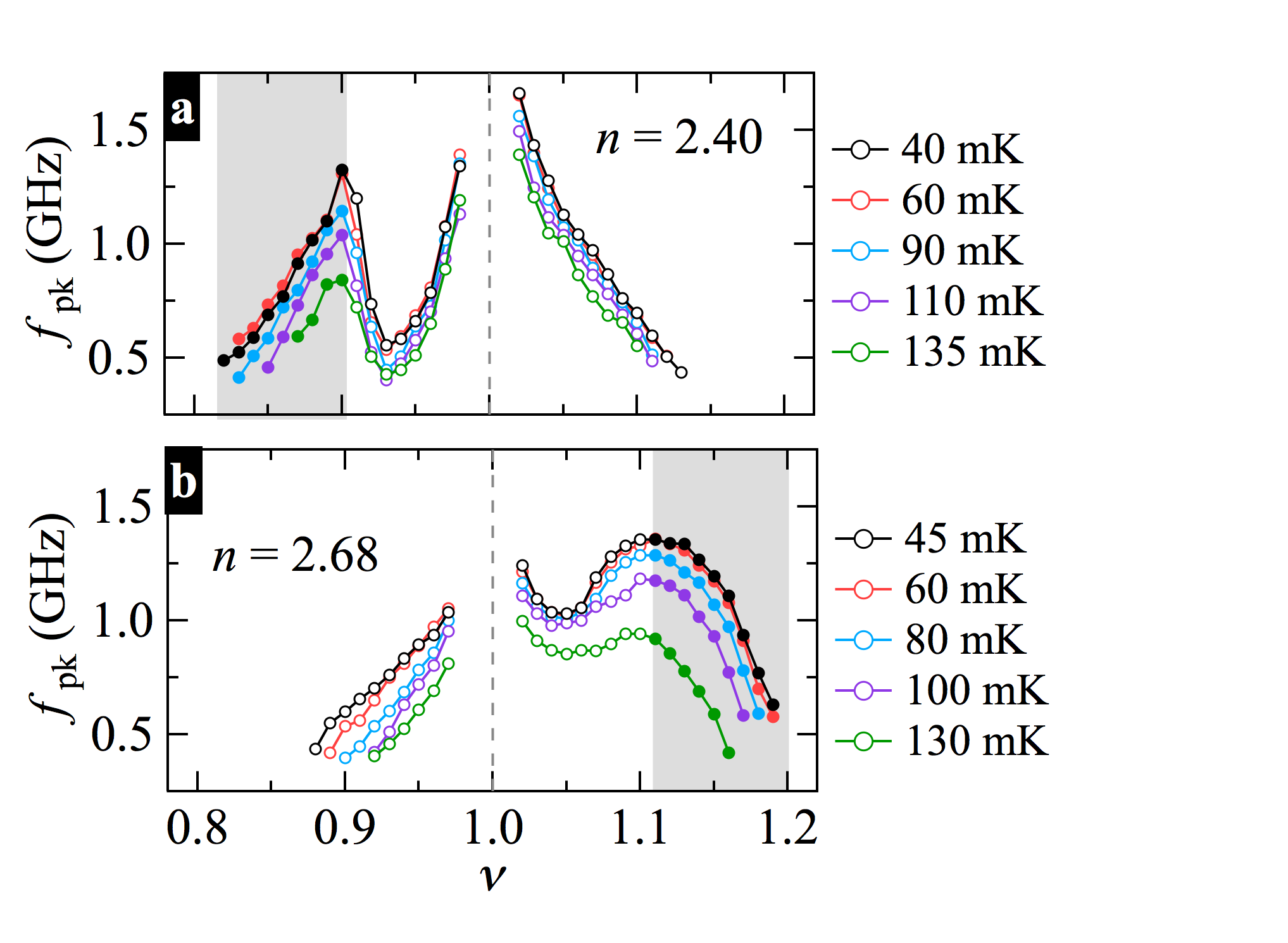}
\vspace{-0.2 in}
\caption{(Color Online)
(a) and (b) $\fpk$ vs $\nu$ for different temperatures, as labeled at right, for $n=2.40$ and $2.68$, respectively.
}
\label{tdepend}
\end{figure}

 \begin{figure}[t]
\includegraphics[width=.4\textwidth]{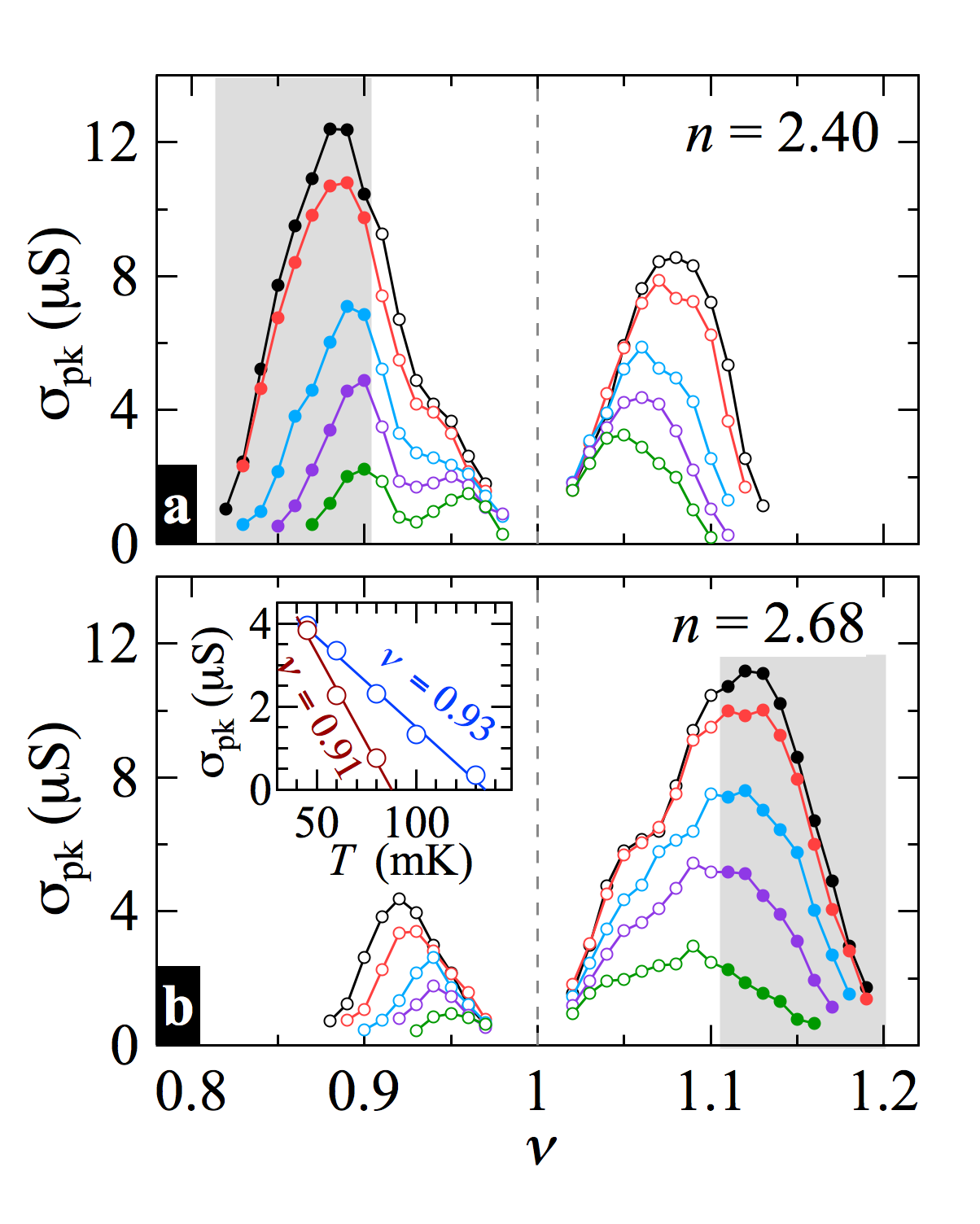}
\vspace{-0.2 in}
\caption{(Color Online)
  Pinning mode resonance maximum   of the real part of the conductivity, $\smax$,   vs $\nu$ at various  temperatures ($T$).  Larger peaks correspond to lower $T$. 
(a)  Data for $n=2.40$, at  $T=40,60,90,110,135$ mK (b) Data for $n=2.68$, at  $T= 45,60,80,100,130$ mK. 
Inset: $\smax$ vs $T$ at $\nu=0.91$ and $\nu=0.93$ for $n=2.68$.  These show  the extrapolation of   $\smax(T)$ to zero, from which we obtain $\tm$.   
}
\label{smaxt}

\includegraphics[width=.4\textwidth]{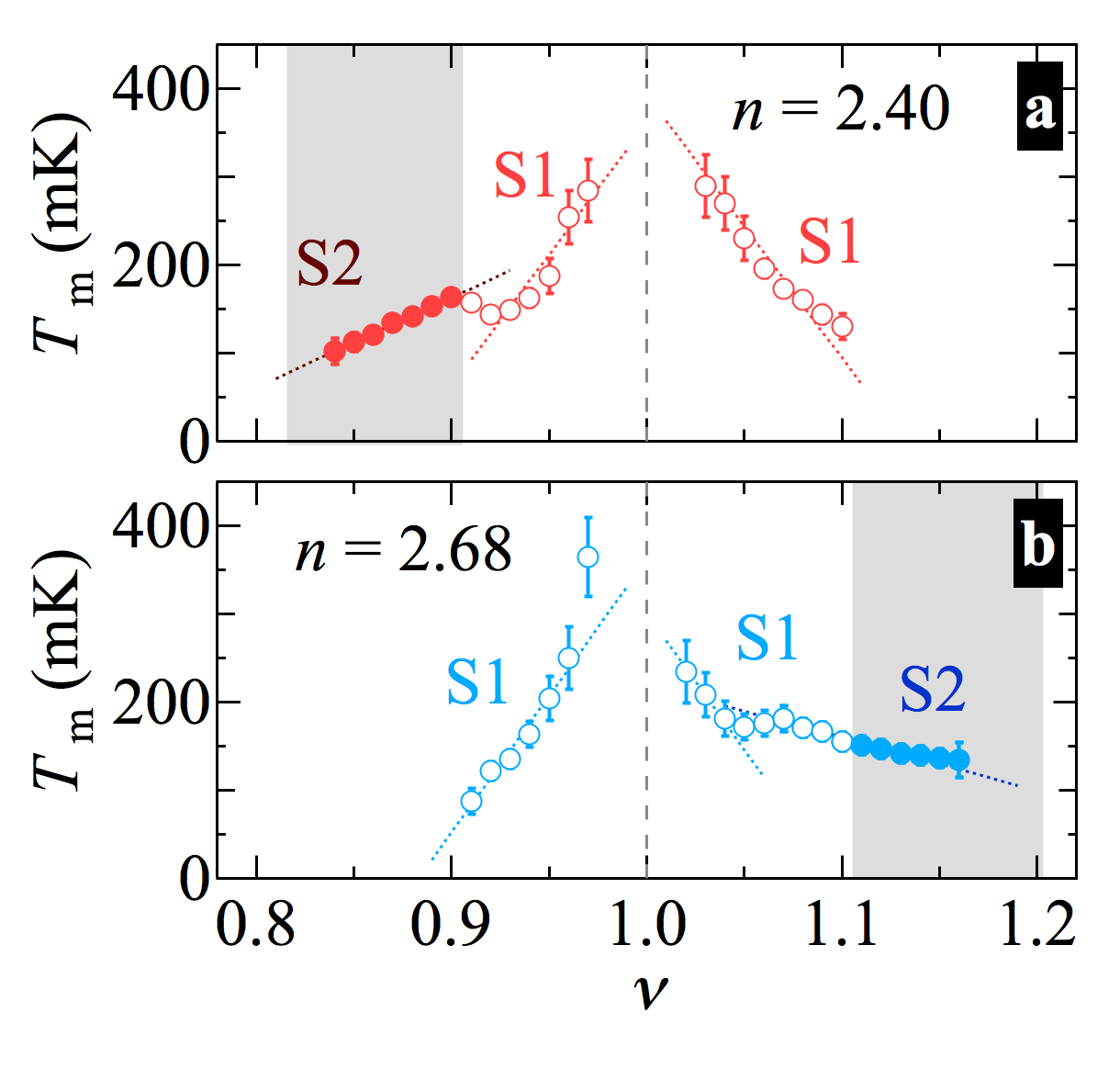}
\vspace{-0.2 in}
\caption{(Color Online)
(a) and (b)  Melting temperature, $\tm$, vs $\nu$ for $n=2.40$ and $n=2.68$, respectively, as obtained from the dependence of the pinning mode on temperature. 
}
\label{tmelt}
\end{figure}

 
\begin{figure}[t]
\includegraphics[width=2.5in]{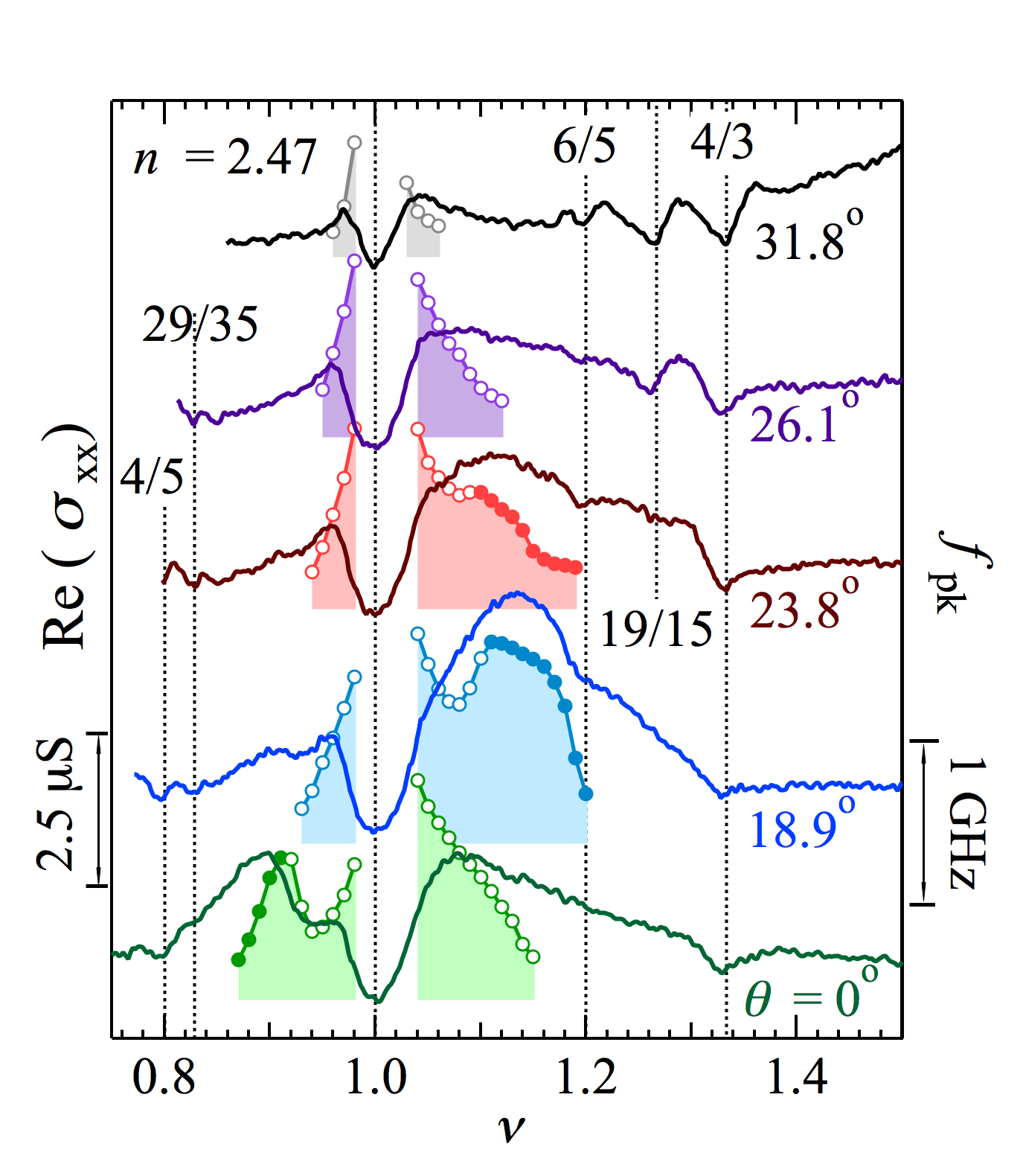}
\vspace{-0.2 in}
\caption{(Color Online)
Real part of the conductivity, Re\,$(\sigma_{\mathrm{xx}})$, obtained at fixed frequency, $f=0.5\,$GHz, vs $\nu$ (left axis) at various $\theta$ as marked.
For each $\theta$, $\fpk$ vs $\nu$ (right axis) shaded to $\fpk=0$.
Traces are vertically offset for clarity.  The vertical lines mark $\nu=4/5$, $1$, $6/5$ and $4/3$ as well as $29/35$ and $19/15$. 
}
\label{n2p47_tilt}
\end{figure}
\subsection{Temperature Dependence}
 
 The interpretation of S1 and S2 as distinct solids is in accord with different temperature ($T$) dependences of their  pinning modes.  
 Figures \,\ref{tdepend}\,(a) and (b) show $\fpk$ vs $\nu$ at various temperatures for $n=2.40$ and $n=2.68$,  with S2 respectively present for $\nu<1$ and $\nu>1$.  The region of well-developed S2, for which \fpk\ decreases with $|\ns|$ is shaded.  As $T$  increases  $\fpk$ decreases, at all $\nu$, indicating the pinning is weaker at higher $T$.  The position of the 
transitions from S1 to S2, as marked by the local maxima in $\fpk$ vs $\nu$, are nearly insensitive to $T$. 

Figures \,\ref{smaxt}\,(a) and (b) show $\smax$  (Re\,$[\sigma_{\mathrm{xx}}(\fpk)]$) vs $\nu$  also for $n=2.40$ and $n=2.68$. The maxima in $\smax$ vs $\nu$ do  not occur at the same place as the maxima in \fpk\ vs $\nu$. 
As $T$ increases, the resonance   $\smax$ decreases 
and the   $\nu$-range of its existence shrinks, with the resonance farthest from $\nu=1$ disappearing first.

Following Ref. \onlinecite{chen:2006}, we estimate  a melting temperature, $\tm$,  as the temperature at which  $\smax$ vs $T$ extrapolates to zero.     The  inset of Fig. \,\ref{smaxt}\,(b) shows  $\smax$ vs $T$ with the lines used to extrapolate, for  $n=2.68$ at $\nu=0.91$  and $\ 0.93$.


 Figures \,\ref{tmelt}\,(a) and  (b) show $\tm$ vs $\nu$ for $n=2.40$ and $n=2.68$, respectively.  As $|\ns|$ increases, the main trend is for $\tm$ to decrease, similar to the 
 behavior seen for $\tm$ vs $\nu$ in Ref. \,\onlinecite{chen:2006}.  
    The  $\nu$-ranges of S2  are marked as shaded regions, and   are identified as explained above from the low-$T$ \fpk\ vs $\nu$  in Fig.\,\ref{tdepend}, as the regions further from $\nu=1$ than the local maximum in \fpk\ vs $\nu$.       Dotted lines, as guides to the eye, are drawn in the S1 and S2 regions of  Figs.\,\ref{tmelt}.   S2 shows  a much weaker dependence of $\tm$ on $\nu$ than S1; the magnitude of the slope for S2 is about a factor of three smaller for $\nu<1$ at $n=2.40$ and a factor of six smaller for $\nu>1$ at $n=2.68$.  In Fig. \,\ref{tmelt}(a ), with  S2  present for $\nu<1$, there is a local maximum   in $\tm$ vs  $\nu$ just  at the  local maximum in \fpk\ vs $\nu$, {\em i.e.} just at the edge of the shaded S2 region.          In Fig.\,\ref{tmelt}(b), with  S2  present for $\nu>1$, the local maximum in  $\tm$ vs $\nu$  occurs at $\nu=1.07$, which is closer to $\nu=1$ than the \fpk\ vs $\nu$ maximum Fig. \,\ref{tdepend}(b), in  the  region of increasing \fpk\ vs $|\ns|$.

\begin{figure*}[t]
\includegraphics[width=.75\textwidth]{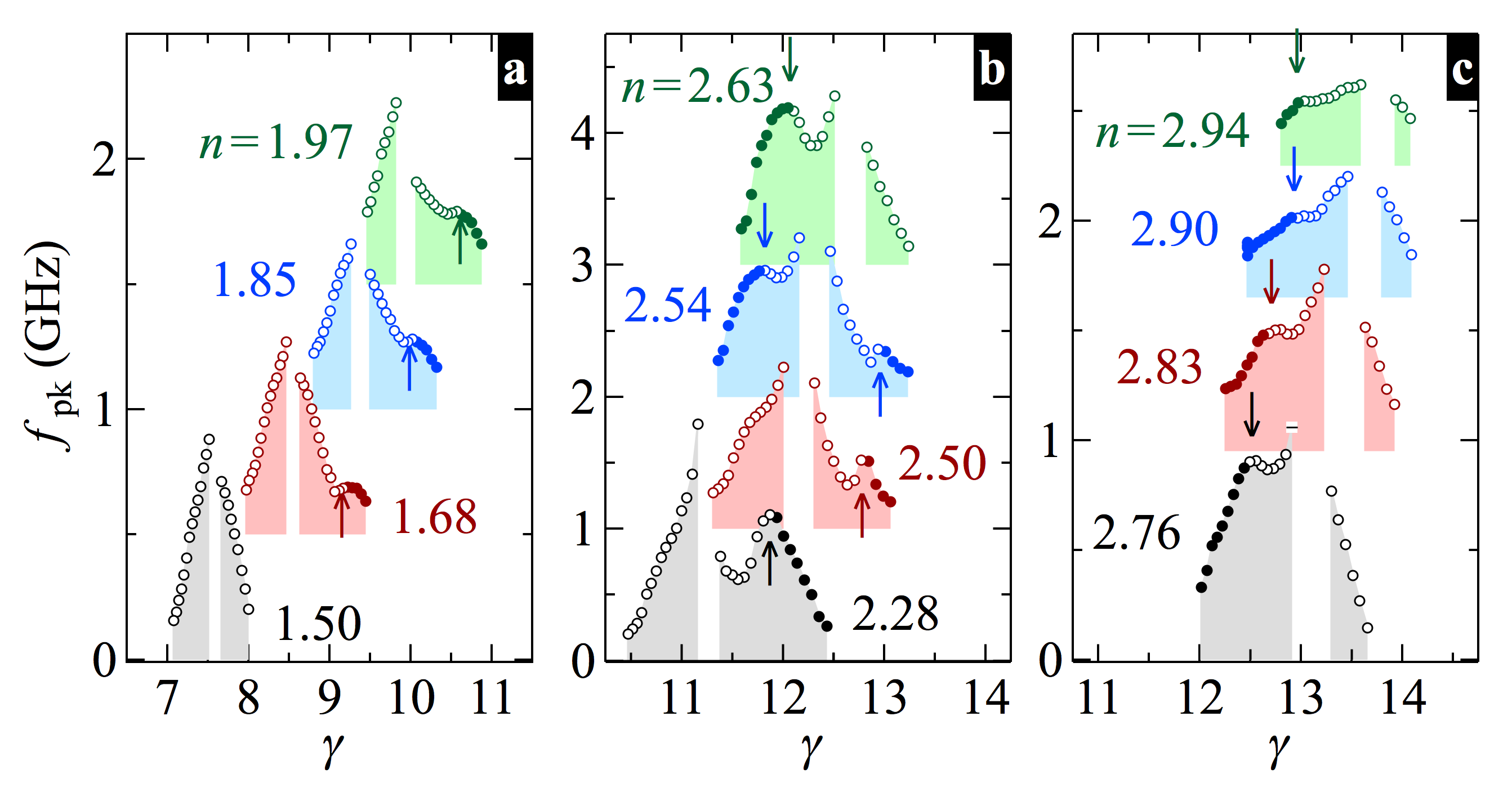}
\vspace{-0.2 in}
\caption{(Color Online)
(a)-(c) $\fpk$ vs $\gamma$ at various $n$, as marked, for $\theta=0\deg$.
Traces are vertically offset from one another for clarity and shaded to $\fpk=0$.
$\uparrow$ and $\downarrow$ mark the S2-S1 transition for $\nu<1$ and $\nu>1$, respectively.
}
\label{gammas}
\end{figure*}
  
\begin{figure}[b]
\includegraphics[width=0.35\textwidth]{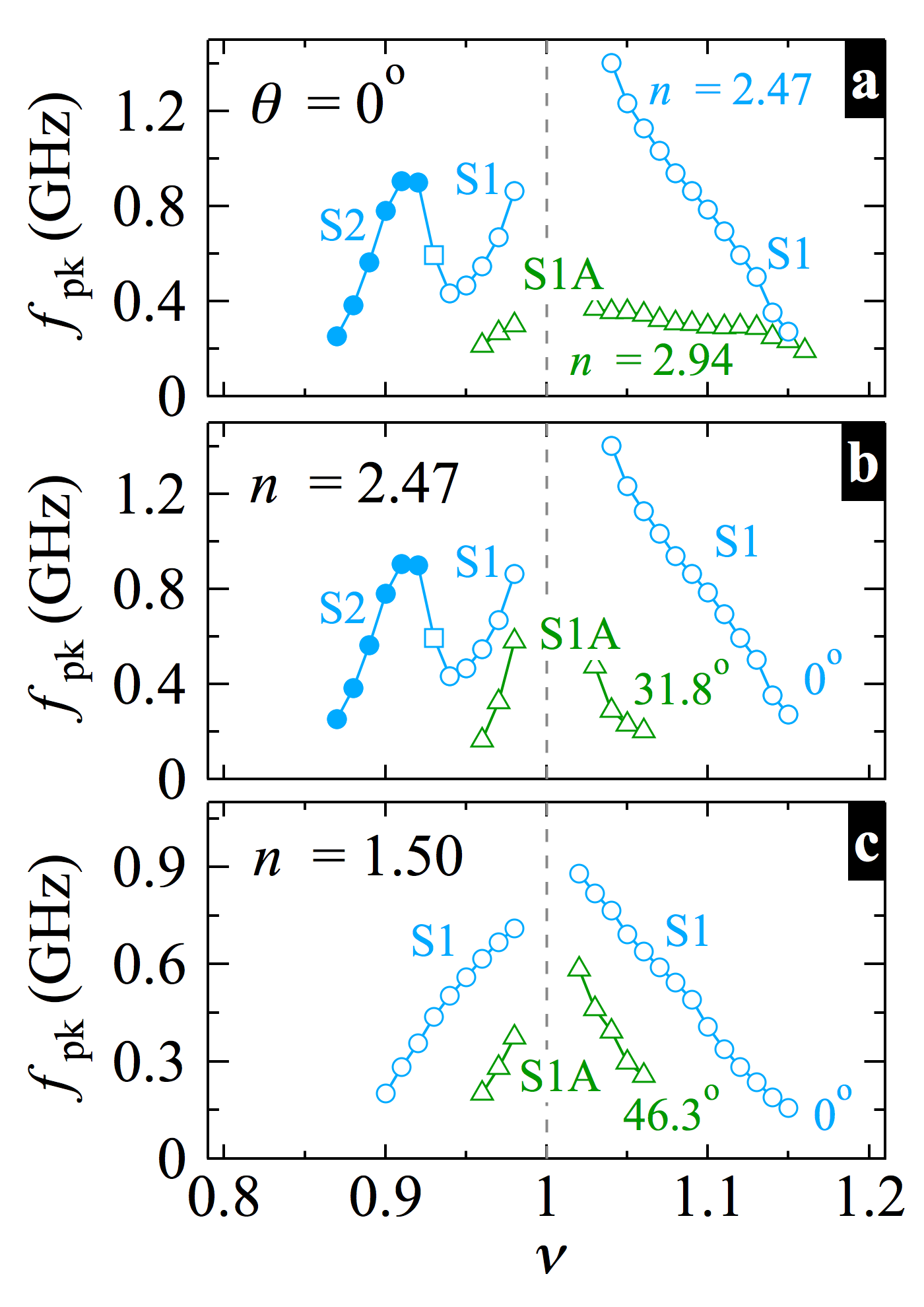}
\vspace{-0.2 in}
\caption{(Color Online) $\fpk$ vs $\nu$. Regions of solids S1, S2 and S1A shown respectively with open circles,  closed circles, and  triangles.
Transition regions between S1 and S2 shown as open squares. 
(a)  $\theta=0\deg$ for $n=2.47$ (upper curves)  and $2.94$ (lower curves), from Sample B. 
(b) $\fpk$ vs $\nu$ at $n=2.47$ for $\theta = 0\deg$ (upper curves) and $31.8\deg$ (lower curves), from Sample B. 
(c) $\fpk$ vs $\nu$ at $n=1.50$ for $\theta = 0\deg$ (upper curves) and $46.3\deg$ (lower curves), from Sample A.  
}
\label{bilayer}
\end{figure}
\subsection{Bilayer Transition}
We have shown that S2 is present for sufficiently high $n$ or $\theta$, but then disappears if $n$ or $\theta$ is increased further, as in Fig.\, \ref{den}(d) for $\nu<1$ and in Fig.\, \ref{tilt}(d).   
It is clear from our data in tilted field and also inferred from dc transport data \cite{liu:2014c}  that larger $\ez$ plays a role in stabilizing S2.  
In this section we show that the disappearance of S2 at still larger $n$ or $\theta$ can be explained by the system undergoing a transition  from a single-layer to  a bilayer state. 
Even when the system  is a bilayer,  the  $\nu=1$ IQHE is clear in the data, and   is understood as a two-component, interlayer-correlated, $\pone$ 
 state,\citep{eisenstein:1992,murphy:1994,lay:1994,lay:1997}  which emerges  when intra- and interlayer interactions have comparable importance.

Single-layer to bilayer transitions are well-known in WQWs  at sufficiently large $n$ or  in-plane magnetic field\citep{suen:1992,suen:1994,manoharan:1996,shayegan:1996,manoharan:1997,lay:1997,hasdemir:2015,mueed:2016}.
The dimensionless parameter $\gamma\equiv\ec/\dsas$, \citep{suen:1994,manoharan:1996,shayegan:1996} is a measure of the tendency of the system to be in a bilayer state, where $\dsas$ is the interlayer tunneling gap. 
Experimentally \citep{suen:1994,manoharan:1996,shayegan:1996} the change from one-component (1C)  to bilayer two-component (2C) states has been shown to occur   when $\gamma\sim 13.5$.    
While $\gamma$ is not well-defined at large in-plane fields, the in-plane field is known to  suppress tunneling and to drive a WQW into a 
bilayer state.\citep{manoharan:1997,lay:1997,hasdemir:2015,mueed:2016}


For $n=2.47$ and several $\theta$, varying through the range where S2 disappears at large tilt,  Fig. \,\ref{n2p47_tilt} shows the 0.5 GHz magneto-conductivity as Re\,$(\sxx)$ vs $\nu$, with $\fpk$ vs $\nu$ superposed.
At $\theta=0\deg$,  Re\,$(\sxx )$ shows   FQHE minima centered at  $\nu=4/5$ and $4/3$, and S2 is present for $\nu<1$ but not for $\nu>1$. 
In the $\theta=18.9\deg$ $\fpk$ vs $\nu$ data in Fig. \,\ref{n2p47_tilt}, S2 is lost for $\nu<1$, but is present for $\nu>1$.    
At the same $\theta$,  a Re\,$(\sxx)$ minimum at $\nu=29/35$    becomes clear. 
Under further sample rotation to $\theta=23.8\deg$, the S2 enhanced-$\fpk$ region for $\nu>1$ is reduced relative to that at $\theta=18.9\deg$, and  with this weakening of S2  a new Re\,$(\sxx)$ minimum at $\nu=19/15$   starts to develop. By $\theta=26.1\deg$ this minimum is well-developed and S2 is no longer observable on either side of $\nu=1$.
Hence minima at $\nu=29/35$ and $19/15$ appear just as S2 fades on their respective sides of $\nu=1$

Transport studies \citep{manoharan:1997} of WQWs have identified the $\nu=29/35$ and $19/15$   FQHE states as inherently 2C and stabilized by a spontaneous  interlayer charge transfer.   For such states, the capacitive energy that works against charge transfer is compensated by the energy gained by forming incompressible FQHE states.
The $\nu=29/35$ state is stabilized by one layer supporting a   $2/5$ FQHE  and the other layer  a $3/7$ FQHE; similarly for $\nu=19/15$ the  layer fillings  are $2/3$ and $3/5$.
Observation of these FQHE states in Re\,$(\sxx)$ vs $\nu$ indicates the system is in a bilayer state. 
The appearance of bilayer-state minima just as S2 disappears    at large $\theta$  suggests  the destabilization of S2 is associated with bilayer formation.


 The   turn-off of S2 at large $n$  and $\theta=0\deg$, at which $\gamma$ is well-defined is demonstrated in  Fig.\,\ref{gammas}, which  shows that S2 fades as $\gamma$ approaches the value for 
the transition to bilayer.  The figure  shows    traces  of $\fpk$ vs $\gamma$ for   various $n$, with the transitions from S1 to S2 marked by arrows.  
Ranges of filling immediately next to   $\nu=1$ correspond  to the gaps between curves for each $n$, and as $n$ is increased, the traces move to higher $\gamma$.   

 In  Fig.\,\ref{gammas}(a) the development of S2 with increasing $n$ occurs as described above, with  $n=1.50$,   S1 only at   low $\gamma$ for which the charge distribution is single-layer. 
On increasing  $n$ to $1.68$, $1.85$ and  $1.97$ the traces    show S2  for $\nu<1$ ($\nu<1$ corresponds to the  high-$\gamma$ side of $\nu=1$) developing and growing stronger.
 Turning to Fig. \ref{gammas}(b), at $n=2.28$,  the \fpk-enhancement of S2 for $\nu<1$ is most pronounced,  
but further increase of $n$ results in a decrease  in the size of this enhancement. 
Increasing  $n$ from $2.54$ to $n=2.63$ causes the enhanced-$\fpk$ region of S2 for $\nu<1$ to disappear  as $\gamma\sim13$ is approached.  For $\nu>1$, on the left sides of the gaps in the curves, S2 appears for $n=2.54$ and strengthens as $n$ goes to $2.63$.  In Fig.\,\ref{gammas}(c), the $\nu<1$ curves     exhibit no sign of S2,  but  S2 for $\nu>1$  also disappears as $\gamma$ approaches $13$. 

Figure  \,\ref{bilayer} compares   \fpk\  vs $\nu$  traces in  single-layer and bilayer states.  
 In Fig.\,\ref{bilayer}(a),  \fpk\  vs $\nu$  is plotted at    $n=2.47$ and $2.94$, for $\theta=0$.  The   $n=2.47$  density is too low to show S2 for $\nu>1$, and shows S1 and S2 for $\nu<1$.   At $n=2.94$, $\nu=1$  has $\gamma=13.7$ and is in the $\pone$ state,   S2  is completely suppressed for $\nu<1$ and nearly suppressed for $\nu>1$.  It is clear that \fpk\ near $\nu=1$ is much smaller  when the 
 $\nu=1$ IQHE is of the $\pone$ type.  We ascribe  this reduced-$\fpk$ resonance  to  a solid of quasi-particles or quasi-holes of the  $\pone$ state, which we refer to as S1A.  Figures\,\ref{bilayer}(b) and (c)  compare \fpk\ vs $\nu$ at $\theta=0$ and at $\theta$  sufficiently large  to suppress S2 on both sides of $\nu=1$.  As in Fig.\,\ref{bilayer}(a),  in the high-$\theta$ cases when $\nu=1$ is bilayer, \fpk\ is markedly reduced from that of S1 (or S2) in the single-layer $\theta=0$ cases, and we ascribe the high-$\theta$ resonance to S1A.   The spectra from which Fig. \ref{bilayer}(c) is derived are shown in Figs. \ref{tilt}(a) and (d).

\section{Discussion}
Earlier microwave studies\cite{hatke:2014} showed that S2 was an extension, as $\nu$ approaches 1, of the RIQHE observed in transport.\citep{liu:2012} 
The RIQHE is insulating as expected for a pinned solid, but does not exhibit a pinning mode resonance except in the range that we denote as supporting S2.  An explanation may be that  the RIQHE is a solid, but has pinning modes that  are overdamped, for example by thermally excited carriers, even at our lowest working temperature.  That would be consistent with the shrinking $\nu$-range of the S2 resonance seen on increasing temperature, as mentioned in section III.B. 
 
Reference \onlinecite{liu:2014c} shows that   the RIQHE emerges  on the high $\ez/\ec$ side  of clear spin transitions exhibited by the 4/5 and 6/5 FQHEs as $\ez/\ec$ was increased.    The high $\ez/\ec$ FQHE states alongside of which the RIQHEs develop are fully spin-polarized, composed of a fully filled $^2$CF $\Lambda$ level from the  partly-filled higher $^2$CF   $\Lambda$ level.  The 4/5 FQHE in this picture is the interacting state of $^2$CFs at their  filling $\nu^{\rm CF}=\nu/(1-2\nu)=-4/3$, with the lower   $\Lambda$ level  fully filled, and the upper level 1/3 filled.   The 6/5 state is explained the same way, taking particle-hole conjugate accounting for spin, $\nu\rightarrow 2-\nu$.   On the high   $\ez/\ec$ side of the spin transition, where the RIQHE and S2  are present,  the upper $\Lambda$ level has the same spin as the lower one.   S2 seen here which lies closer to $\nu=1$ (which is  $\nu^{\rm CF}=-1$)    than the $4/5$ or $6/5$ FQHEs  is inferred also to be composed of CFs from the same $\Lambda$ levels, but with the upper $\Lambda$ level  {\em less filled}.  The picture advanced in Ref. \onlinecite{liu:2014c} explicitly requires interaction of CFs, as do CF WSs.   The present $n$ and $\theta$  dependence studies on either side of $\nu=1$  show   directly from the pinning mode that S2 requires a sufficiently large $\ez/\ec$, consistent with the interpretation of S2 as composed of fully spin-polarized states.  
%

If the lower  $\Lambda$ level remains filled, S2 can then be thought of as composed of carriers in the partly-filled upper   $\Lambda$ level.  
 A different case of CF solid in the presence of CF liquid, quasi-particles of an FQHE liquid coexisting with that liquid, was described theoretically in Refs.\,\onlinecite{archer:2013} and \onlinecite{archer:2011}, where it was referred to as type-2 CF WS.    Such a WS, of quasi-particles and -holes of the 1/3 FQHE state, was reported from microwave spectroscopy.\cite{zhu:2010b}\ \ 
(Type-1 CF WS on the other hand referred to WSs without the coexisting liquid, such as can be found at the low-$\nu$ termination of the FQHE series.)  Solids of type 2 were shown to have different wave functions than those of type I.   
 The different \fpk\ and $\tm$  seen here for  S1 and S2 raise the question of whether the underlying filled  $\Lambda$ level present for S2 is affecting its wave functions or playing a role in stabilizing it when both $\Lambda$ levels have the same spin.

The different $\nu$-dependences of  $ \tm$ for S1 and S2, as shown in  Fig.\,\ref{tmelt},  confirm our interpretation of S1 and S2 as distinct phases.
In weak pinning,\cite{chitra:2001,fertig:1999,fogler:2000}  enhanced $\fpk$ (as for S2)  is an indication of a smaller shear modulus (as in the density effect), or an increased ``effective  disorder".  Effective  disorder    is the  disorder  potential integrated over the charge profile for a carrier.  Studies\citep{chen:2006} of pinned WSs for $\nu$ well below the low-$\nu$ termination of the FQHE series show that the pinning affects $\tm$, such that larger disorder produces both larger $\fpk$ and higher $\tm$ for the same $\nu$, while $n$ (which affects the shear modulus) had little effect on $\tm$ when $\nu$ was held constant in a given sample.   For S2, $\tm$ and $\fpk$ are enhanced relative to their values for S1  extrapolated to the larger $|\ns|$ at which S2 exists,  suggesting  S2 experiences larger effective disorder.

Regardless of $\ez/\ec$, when $|\ns|$ is below about $0.09$,  S2 does not survive, and S1 prevails.  
In theories\cite{archer:2011,archer:2013,rhim:2015} of single-layer systems, recently corroborated near $\nu=1$ in tunneling experiments,\cite{jang:2017} low $|\ns|$ favors CF crystals of larger vortex number $2p$. 
Applying those theories to the S1 quasi-particles and -holes near $\nu=1$, in the range of partial filling $|\ns|$ below 0.09,  $2p\ge 6$.
However, it is not unreasonable to assume that the $|\ns|$-range of $2p=4$ may be modified when the theory is applied to the case of the IQHE instead of the low-$\nu$ case.

Neglecting the correlations that give rise to CFs, Skyrme crystals  have also been reported\cite{brey:1995,bayot:1996,cote:1997,zhu:2010b} near $\nu=1$.    
Reference\,\onlinecite{zhu:2010b} reported $\fpk$ to be reduced when $|\ns|$ was small and for low $\ez/\ec$, {\em i.e.} under conditions which Ref.\,\onlinecite{cote:1997} discussed  crystallization of spin skyrmions containing two or more spins.  The reduction of \fpk\ in  Ref.\,\onlinecite{zhu:2010b} was strongest for the resonances seen at the smallest $|\ns|$, and was strongly $\theta$-dependent for samples of similar $n$ to those in this work.  This \fpk\ reduction  is clearly not observed here, for example in Fig.\,\ref{fpkt}(a), in which the points of smallest 
$|\ns|$, in S1, have little $\theta$ dependence.     The present QW samples,  unlike the samples of Ref.\,\onlinecite{zhu:2010b}, are wide enough  and have sufficiently high $n$ that two subbands are occupied at $B_t=0$.  It may be that the correlations of CFs
overwhelm the spin skyrmion effects    in the present case. 

S1A, which supplants S1 in the extreme case of our measurements for large $\ez$ or  $n$, is taken as a WS composed of excitations  of the bilayer  $\pone$ state.   It has been proposed\cite{girmacperspec,moon:1995} that such excitations, for the layer separations and $\Delta_{\rm SAS}$ of the present samples, are merons connected together by a domain wall.  A variety of bilayer WS phases near $\nu=1$ were considered in Ref. \onlinecite{bourassa:2006}, although those calculations apply at smaller layer separations than  those of the present samples.

\section{Summary}
To summarize, we have studied systematically the evolution with $n$ and $\theta$ of the solids S1 and S2 found near $\nu=1$  in WQWs. 
We find that S2  is an extension of the RIQHE of Refs.\,\onlinecite{liu:2012} and \onlinecite{liu:2014c} to $\nu$ closer to 1, at which a pinning mode can be observed.   
The appearance of S2 as $n$ or $\theta$ increase is enabled by sufficiently large Zeeman energy, $\ez/\ec$ above 0.017, to produce a fully-polarized combined state of $\Lambda$ levels.  
S2 however, does not extend closer to $\nu=1$ than 0.09.   
At  larger $n$ or $\theta$,  S2 disappears as the state becomes a bilayer.  
We also have seen pinning modes due to excitations of the $\pone$ state for $\nu=1$, and find they have smaller $\fpk$ than for S1 in single-layer states.

\section{Acknowledgements}
The microwave spectroscopy work at NHMFL was supported through Department of Energy Basic Energy Sciences grant DE-FG02-05-ER46212 at NHMFL/FSU.   
The National High Magnetic Field Laboratory (NHMFL),  is supported by NSF Cooperative Agreement No. DMR-0654118, by the State of Florida, and by the DOE.
The work at Princeton University was supported by the Department of Energy Basic Energy Sciences (Grant No. DE-FG02-00-ER45841) for characterization, and the National Science Foundation (Grants No. DMR 1709076 and MRSEC DMR 1420541), and the Gordon and Betty Moore Foundation (Grant No. GBMF4420) for sample fabrication.


\end{document}